\documentclass[conference]{IEEEtran}
\usepackage{dsfont,graphicx,color,epsf,epsfig,verbatim,amssymb,amsmath,amsthm}
\usepackage{latexsym,amsfonts,caption,array,cite,subfigure,multicol,multirow}

\newcommand{\argmax}{\operatornamewithlimits{argmax}}

\setkeys{Gin}{draft=false}

\newtheorem{lemma}{Lemma}[section]

\newtheorem{theorem}{Theorem}[section]

\newcommand{\G}{\mathbf{G}}

\IEEEoverridecommandlockouts
\begin{document}
\title{Polar Codes for Some Multi-terminal Communications Problems}% can use linebreaks \\
\author{\IEEEauthorblockN{Aria G. Sahebi and S. Sandeep Pradhan  \thanks{This work was supported by NSF grant CCF-1116021.}}
\IEEEauthorblockA{Department of Electrical Engineering and Computer Science,\\
University of Michigan, Ann Arbor, MI 48109, USA.\\
Email: \tt\small ariaghs@umich.edu, pradhanv@umich.edu}}

%\markboth{Draft}
%{Sahebi \MakeLowercase{\textit{et al.}}: Multilevel Polarization of Polar Codes Over Arbitrary Discrete Memoryless Channels}

\maketitle

\begin{abstract}
It is shown that polar coding schemes achieve the known achievable rate regions for several multi-terminal communications problems including lossy distributed source coding, multiple access channels and multiple descriptions coding. The results are valid for arbitrary alphabet sizes (binary or non-binary) and arbitrary distributions (symmetric or asymmetric).
\end{abstract}

\section{Introduction}
Polar codes were recently proposed by Arikan \cite{arikan_polar} to achieve the symmetric capacity of binary input channels. This result was later generalized to arbitrary discrete memoryless channels \cite{sasoglu_polar_q,mori_polar,sahebi_multilevel_polar_ieee,Park_Barg_Polar}. Polar coding schemes were also developed to achieve the symmetric rate-distortion function for arbitrary discrete memoryless sources \cite{korada_lossy_source,karzand_polar_source,sahebi_polar_source}. Polar coding results for asymmetric cases are developed in \cite{honda_polar_extension}. Among the existing works on the application of polar codes for multi-terminal cases we note \cite{arikan_source_polarization,arikan_slepian_wolf,korada_lossy_source,abbe_randomness_extraction} for distributed source coding, \cite{sasoglu_MAC,Abbe_Polar_Mac} for the multiple access channels and \cite{goela_polar_broadcast_arxiv} for broadcast channels.\\

In \cite{sahebi_nested_polar_arxiv}, it is shown that nested polar codes can be used to achieve the Shannon capacity of arbitrary discrete memoryless channels and the Shannon rate-distortion function for discrete memoryless sources. In this paper, we show that nested polar codes can achieve the best known achievable rate regions for several multi-terminal communication systems. We present several examples in this paper, including the distributed source coding problem, multiple access channels, computation over MAC, broadcast channels and multiple description coding to illustrate how these codes can be employed to have an optimal performance for multi-terminal cases. The results of this paper are general regarding the size of alphabets (binary or non-binary) using the approach of \cite{sahebi_multilevel_polar_ieee}. The special case where the alphabets are binary is discussed in \cite{sahebi_nested_polar_arxiv} for the lossy source coding problem. In addition, the results of this paper are general regarding the distributions i.e., we do not assume uniform distributions on the channel inputs or the source alphabets.

This paper is organized as follows: In Section \ref{prel} we state some preliminaries. In Section \ref{section:BT}, we consider the distributed source coding problem and show that polar codes achieve the Berger-Tung rate region. In Section \ref{section:KM}, we consider a distributed source coding problem in which the decoder is interested in decoding the sum of auxiliary random variables and show that polar codes have the optimal performance (this scheme has the Korner-Marton scheme as a special case). In Section \ref{section:MAC}, we show that polar codes achieve the capacity region for multiple access channels. In Section \ref{section:Computation}, we show that polar codes have an optimal performance for the problem of computation over MAC where the decoder is interested in the sum of variables. In Section \ref{section:BC}, we study the performance of polar codes for broadcast channels. In Section \ref{section:MD}, we show that polar codes are optimal for the multiple description problem. Finally, in Section \ref{section:discussion}, we discuss briefly other possible problems and extensions to multiple user (more that two) cases.
\section{Preliminaries} \label{prel}

\subsubsection{Channel Parameters}
For a channel $(\mathcal{X},\mathcal{Y},W)$, assume $\mathcal{X}$ is equipped with the structure of a group $(\G,+)$. The symmetric capacity is defined as $\bar{I}(W)=I(X;Y)$ where the channel input $X$ is uniformly distributed over $\mathcal{X}$ and $Y$ is the output of the channel. For $d\in\G$, we define
\begin{align*}%\label{eqn:Zd}
Z_d(W)=\frac{1}{q}\sum_{x\in \G}\sum_{y\in \mathcal{Y}}\sqrt{W(y|x)W(y|x+d)}
\end{align*}
and for $H\le \G$ define $Z^H(W)=\sum_{d\notin H} Z_d(W)$.
%; i.e. for $q=|\mathcal{X}|$,
%%\begin{align*}
%%I^0(W)=\sum_{x\in\mathcal{X}}\sum_{y\in\mathcal{Y}}\frac{1}{q}W(y|x)\log\frac{W(y|x)}{\displaystyle\sum_{{\tilde{x} \in\mathcal{X}}}\frac{1}{q}W(y|\tilde{x})}
%%\end{align*}
%The Bhattacharyya distance between two distinct input symbols $x$ and $\tilde{x}$ is defined as
%\begin{align*}
%Z(W_{\{x,\tilde{x}\}})=\sum_{y\in\mathcal{Y}}\sqrt{W(y|x)W(y|\tilde{x})}
%\end{align*}
%and the average Bhattacharyya distance is defined as
%\begin{align*}
%Z(W)=\sum_{\substack{x,\tilde{x}\in \mathcal{X}\\x\ne\tilde{x}}}\frac{1}{q(q-1)}Z(W_{\{x,\tilde{x}\}})
%\end{align*}
%where $q=|\mathcal{X}|$. We use the following two quantities in the paper extensively:
%\begin{align*}
%&D_d(W)=\frac{1}{2q}\sum_{u\in \mathcal{U}} \sum_{x\in\mathcal{X}} \left|W(x|u)-W(x|u+d)\right|\\
%&\tilde{D}_d(W)=\frac{1}{2q}\sum_{u\in \mathcal{U}} \sum_{x\in\mathcal{X}} \left(W(x|u)-W(x|u+d)\right)^2
%\end{align*}
%where $d$ is some element of $\G$ and $+$ is the group operation.\\

\subsubsection{Binary Polar Codes}
For any $N=2^n$, a polar code of length $N$ designed for the channel $(\mathds{Z}_2,\mathcal{Y},W)$ is a linear (coset) code characterized by a generator matrix $G_N$ and a set of indices $A\subseteq \{1,\cdots,N\}$ of \emph{almost perfect channels}. The set $A$ is a function of the channel. The decoding algorithm for polar codes is a specific form of successive cancelation \cite{arikan_polar}.\\% The generator matrix for polar codes is defined as $G_N=B_NF^{\otimes n}$ where $B_N$ is a permutation of rows, $F=\left[ \begin{array}{cc}1 & 0\\1 & 1\end{array} \right]$ and $\otimes$ denotes the Kronecker product.

\subsubsection{Polar Codes Over Abelian Groups}
For any discrete memoryless channel, there always exists an {Abelian group} of the same size as that of the channel input alphabet. Polar codes for arbitrary discrete memoryless channels (over arbitrary Abelian groups) are introduced in \cite{sahebi_multilevel_polar_ieee}. For various notations used in this paper, we refer the reader to \cite{sahebi_multilevel_polar_ieee} and \cite{sahebi_nested_polar_arxiv}.\\

%\subsubsection{Group Codes}
%Let the channel input alphabet $\mathcal{X}$ be equipped with the structure of a finite Abelian group $\G$ of the same size. Then the channel is specified by $(\G,\mathcal{Y},W)$. A group code over $\G$ of length $N$ for this channel is any {subgroup} of $\G^N$. The group capacity of a channel $(\G,\mathcal{Y},W)$ is the maximum achievable rate using group codes over $\G$ for this channel. Group codes generalize the notion of linear codes over {fields} to channels with composite input alphabet sizes. A coset code is a shift of a group code by a constant vector.\\

%\subsubsection{Notation}%We say a function $f:\mathds{R}\rightarrow\mathds{R}$ is $O(\epsilon)$ if it's right limit is zero at zero.
%We denote by $O(\epsilon)$ any function of $\epsilon$ which is right-continuous around $0$ and that $O(\epsilon)\rightarrow 0$ as $\epsilon\downarrow 0$.\\
%For positive integers $N$ and $r$, let $\{A_0,A_1,\cdots,A_r\}$ be a partition of the index set $\{1,2,\cdots,N\}$. Given sets $T_t$ for $t=0,\cdots,r$, the direct sum $\bigoplus_{t=0}^r T_t^{A_t}$ is defined as the set of all tuples $u_1^N=(u_1,\cdots,u_N)$ such that $u_i\in T_t$ whenever $i\in A_t$.\\

\section{Distributed Source Coding: The Berger-Tung Problem}\label{section:BT}
In the distributed source coding problem, two separate sources $X$ and $Y$ communicates with a centralized decoder. Let $\mathcal{X},\mathcal{Y}$ and $\mathcal{U},\mathcal{V}$ be the source and the reconstruction alphabets of the two terminals and assume $X$ and $Y$ have the joint distribution $p_{XY}$. Let $d_1:\mathcal{X}\times\mathcal{U}\rightarrow \mathds{R}^+$ and $d_2:\mathcal{Y}\times\mathcal{V}\rightarrow \mathds{R}^+$ be the distortion measures for terminals $X$ and $Y$ respectively. We denote this source by $(\mathcal{X},\mathcal{Y},\mathcal{U},\mathcal{V},p_{XY},d_1,d_2)$. Let $U$ and $V$ be auxiliary random variables taking values from $\mathcal{U}$ and $\mathcal{V}$ respectively such that $U\leftrightarrow X\leftrightarrow Y\leftrightarrow V$, $\mathds{E}\{d_1(X,U)\}\le D_1$ and $\mathds{E}\{d_2(Y,V)\}\le D_2$ for some distortion levels $D_1,D_2\in\mathds{R}^+$. It is known by the Berger-Tung coding scheme that the tuple $(R_1,R_2,D_1,D_2)$ is achievable if $R_1\ge I(X;U)-I(U;V)$, $R_2\ge I(Y;V)-I(U;V)$ and $R_1+R_2\ge I(X;U)+I(Y;V)-I(U;V)$. In this section, we prove the following theorem:
\begin{theorem}
For a source $(\mathcal{X},\mathcal{Y},\mathcal{U},\mathcal{V},p_{XY},d_1,d_2)$, assume $\mathcal{U}$ and $\mathcal{V}$ are finite. Then the Berger-Tung rate region is achievable using nested polar codes.% \eqref{eqn:Shannon_C}.
\end{theorem}

It suffices to show that the rates $R_1=I(X;U)-I(U;V)$ and $R_2=I(Y;V)$ achievable. Let $\G$ be an Abelian group of the size larger than or equal to the size of both $\mathcal{U}$ and $\mathcal{V}$. Note that for the source $Y$, we can use a nested polar codes as introduced in \cite{sahebi_nested_polar_arxiv} to achieve the rate $I(Y;V)$. Furthermore, we have access to the outcome $v_1^N$ of $V_1^N$ at the decoder with high probability. It remains to show that the rate $R_1=I(X;U)-I(U;V)$ is achievable when the sequence $v_1^N$ with $d_2(y_1^N,v_1^N)\le D_2$ is available at the decoder.\\%\in\G^N

\begin{figure}[!h]
\begin{minipage}[b]{0.23\textwidth}
\centering
\includegraphics[scale=.8]{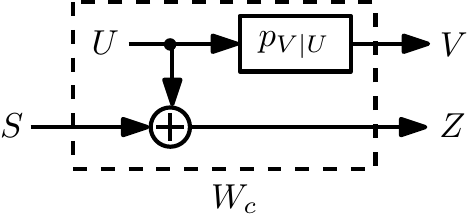}
\caption{\small Test channel for the inner code (the channel coding component)}
\label{fig:Wc_RD}
\end{minipage}
%\hspace{0.5cm}
\begin{minipage}[b]{0.23\textwidth}
\centering
\includegraphics[scale=.8]{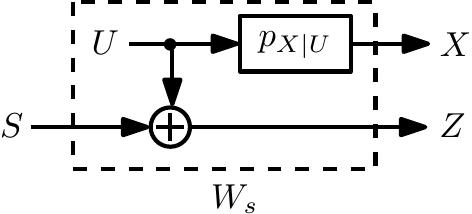}
\caption{\small Test channel for the outer code (the source coding component)}
\label{fig:Ws_RD}
\end{minipage}
\end{figure}

Given the test channel $p_{X|U}$, define the artificial channels $(\G,\G^2,W_c)$ and $(\G,\mathcal{X}\times\G,W_s)$ such that for $s,z\in \G$ and $x\in \mathcal{X}$, $W_c(v,z|s)=p_{VU}(v,z-s)$ and $\quad W_s(x,z|s)=p_{XU}(x,z-s)$. These channels have been depicted in Figures \ref{fig:Wc_RD} and \ref{fig:Ws_RD}.
Let $S$ be a random variable uniformly distributed over $\G$ which is independent from $X$ and $U$. It is straightforward to show that in this case, $Z$ is also uniformly distributed over $\G$. Similarly to the point-to-point result \cite{sahebi_nested_polar_arxiv}, we can show that the symmetric capacities of the channels $W_c$ and $W_s$ are given by $\bar{I}(W_c)=\log q-H(U|V)$ and $\bar{I}(W_s)=\log q-H(U|X)$. We employ a nested polar code in which the inner code is a good channel code for the channel $W_c$ and the outer code is a good source code for $W_s$. The rate of this code is equal to $R=\bar{I}(W_s)-\bar{I}(W_c)=I(X;U)-I(U;V)$. The rest of this section is devoted to some general definitions and lemmas which are used in the proofs.
\begin{lemma}\label{lemma:Wc_deg_Ws_DSC2}
The channel $W_c$ is stochastically degraded with respect to the channel $W_s$.
\end{lemma}
\begin{IEEEproof}
In the Definition \cite[Definition III.1]{sahebi_nested_polar_arxiv}, let the channel $(\mathcal{X}\times \G,\G^2,W)$ be such that for $v,z,z'\in\G$ and $x\in\mathcal{X}$, $W(v,z|x,z')=p_{V|X}(v|x) \mathds{1}_{\{z=z'\}}$.
%Then for $s,z\in\G$,
%\begin{align*}
%\sum_{\substack{z'\in\G\\x\in\mathcal{X}}} W_s(x,z'|s) p_{V|X}(v|x) \mathds{1}_{\{z=z'\}}
%&= \sum_{\substack{z'\in\G\\x\in\mathcal{X}}} P_{XU}(x,z'-s)\cdot p_{V|X}(v|x)\mathds{1}_{\{z=z'\}}\\
%&= \sum_{x\in\mathcal{X}} P_{XU}(x,z-s)\cdot p_{V|X}(v|x)\\
%&\stackrel{(a)}{=} \sum_{x\in\mathcal{X}} P_{XU}(x,v,z-s)\\
%&= P_{VU}(v,z-s) = W_c(z|s)
%\end{align*}
%where $(a)$ follows since the Markov chain $U\leftrightarrow X\leftrightarrow Y\leftrightarrow V$ holds.
\end{IEEEproof}

Let $N=2^n$ for some positive integer $n$ and let $G$ be the corresponding $N\times N$ generator matrix for polar codes. For $i=1,\cdots,N$, and for $z_1^N,a_1^N\in \G^N$, $v_1^N\in\mathcal{V}^N$ and $x_1^N\in\mathcal{X}^N$, let
\begin{align*}
&W_{c,N}^{(i)}(z_1^n,v_1^N,a_1^{i-1}|a_i)=\sum_{a_{i+1}^N\in \G^{N-i}} \frac{1}{q^{N-1}} W_c^N(z_1^N,v_1^N|a_1^NG)\\% = \sum_{a_{i+1}^N\in \G^{N-i}} \frac{1}{q^{N-1}} p_{VU}^N(v_1^N,z_1^N-a_1^NG)\\
&W_{s,N}^{(i)}(x_1^N,z_1^n,a_1^{i-1}|a_i)=\sum_{a_{i+1}^N\in \G^{N-i}} \frac{1}{q^{N-1}} W_s^N(x_1^N,z_1^N|a_1^NG)\\% = \sum_{a_{i+1}^N\in \G^{N-i}} \frac{1}{q^{N-1}} p_{XU}^N(x_1^N,z_1^N-a_1^NG)\\
\end{align*}
%Let $J_n$ be a uniform random variable over the set $\{1,2,\cdots,N\}$ and define
%\begin{align*}
%&I^n(W_c)=\bar{I}(W_{c,N}^{(J_n)})\\
%&I^n(W_s)=\bar{I}(W_{s,N}^{(J_n)})\\
%&Z^n(W_c)=Z(W_{c,N}^{(J_n)})\\
%&Z^n(W_s)=Z(W_{s,N}^{(J_n)})
%\end{align*}

Let the random vectors $X_1^N,Y_1^N,U_1^N,V_1^N$ be distributed according to $P_{XYUV}^N$ and let $Z_1^N$ be a random variable uniformly distributed over $\G^N$ which is independent of $X_1^N,Y_1^N,U_1^N,V_1^N$. Let $S_1^N=Z_1^N-U_1^N$ and $A_1^N=S_1^N G^{-1}$ (Here, $G^{-1}$ is the inverse of the mapping $G:\G^N\rightarrow \G^N$). In other words, the joint distribution of the random vectors is given by
\begin{align*}
&p_{A_1^NS_1^NU_1^NV_1^NX_1^NZ_1^N}(a_1^N,s_1^N,u_1^N,v_1^N,x_1^N,z_1^N)\\
&=\frac{1}{q^N} p_{XUV}^N(x_1^N,u_1^N,v_1^N) \mathds{1}_{\{s_1^N=a_1^NG,u_1^n=z_1^N-a_1^NG\}}
\end{align*}
%This implies
%\begin{align*}
%p_{A_1^NV_1^NX_1^NZ_1^N}(a_1^Nv_1^N,x_1^N,z_1^N)&=\frac{1}{q^N} p_{XUV}^N(x_1^N,z_1^N-a_1^NG,v_1^N)\\
%p_{A_1^NV_1^NZ_1^N}(a_1^N,v_1^N,z_1^N)&=\frac{1}{q^N} p_{UV}^N(z_1^N-a_1^NG,v_1^N)
%\end{align*}

\subsection{Sketch of the proof}
%\subsection{Review of Polar Codes for Arbitrary Sources and Channels}
The following theorems from \cite{sahebi_multilevel_polar_ieee} state the standard channel coding and source coding polarization phenomenons for the general case.
\begin{theorem}\label{theorem:polar_channelG}
For any $\epsilon>0$ and $0< \beta <\frac{1}{2}$, there exist a large $N=2^n$ and a partition $\{A_H|H\le \G\}$ of $[1,N]$ such that for $H\le \G$ and $i\in A_H$, $\left|\bar{I}(W_{c,N}^{(i)}) -\log \frac{|\G|}{|H|} \right|<\epsilon$ and $Z^H(W_{c,N}^{(i)})< 2^{-N^{\beta}}$. Moreover, as $\epsilon\rightarrow 0$ (and $N\rightarrow \infty$), $\frac{|A_H|}{N}\rightarrow p_H$ for some probabilities $p_H,H\le \G$ adding up to one with $\sum_{H\le \G} p_H \log \frac{|\G|}{|H|}=\bar{I}(W_c)$.
\end{theorem}

\begin{theorem}\label{theorem:polar_sourceG}
For any $\epsilon>0$ and $0< \beta <\frac{1}{2}$, there exist a large $N=2^n$ and a partition $\{B_H|H\le \G\}$ of $[1,N]$ such that for $H\le \G$ and $i\in B_H$, $\left|\bar{I}(W_{s,N}^{(i)}) -\log \frac{|\G|}{|H|} \right|<\epsilon$ and $Z^H(W_{s,N}^{(i)})< 2^{-N^{\beta}}$. Moreover, as $\epsilon\rightarrow 0$ (and $N\rightarrow \infty$), $\frac{|B_H|}{N}\rightarrow q_H$ for some probabilities $q_H,H\le \G$ adding up to one with $\sum_{H\le \G} q_H \log \frac{|\G|}{|H|}=\bar{I}(W_s)$.
\end{theorem}

%\begin{lemma}
%If the channel $(\G,\mathcal{Y}_1,W_1)$ is degraded with respect to the channel $(\G,\mathcal{Y}_2,W_2)$ in the sense of Definition (?), then for any $d\in G$,
%\begin{align*}
%Z_d(W_1)\ge Z_d(W_2)
%\end{align*}
%\end{lemma}
%\begin{IEEEproof}
%Follows using the Cauchy-Schwarz inequality in a straightforward fashion.
%\end{IEEEproof}
%
%\begin{lemma}\label{lemma:Zdc_Zds_RDG}
%For $i=1,\cdots,N$ and for $d\in \G$ and $H\le\G$, $Z_d(W_{c,N}^{(i)})\ge Z_d(W_{s,N}^{(i)})$ and $Z^H(W_{c,N}^{(i)})\ge Z^H(W_{s,N}^{(i)})$.
%\end{lemma}
%\begin{IEEEproof}
%Follows from Lemma \ref{lemma:Wc_deg_Ws_RD2}, Lemma[???] and Lemma[???].
%\end{IEEEproof}

For $H\le \G$, define
\begin{align*}
&A_H=\Big\{i\in[1,N]\Big|Z^H(W_{c,N}^{(i)})<2^{-N^{\beta}},\\
&\qquad\qquad\qquad \nexists K\le H: Z^K(W_{c,N}^{(i)})<2^{-N^{\beta}}\Big\}\\
&B_{H}=\Big\{i\in[1,N]\Big|Z^{H}(W_{s,N}^{(i)})<1-2^{-N^{\beta}},\\
&\qquad\qquad\qquad \nexists K\le H: Z^{K}(W_{c,N}^{(i)})<1-2^{-N^{\beta}}\Big\}
%&A_H=\left\{i\in[1,N]\left|Z^H(W_{c,N}^{(i)})<2^{-N^{\beta}}, \forall K\le H: Z^K(W_{c,N}^{(i)})>2^{-N^{\beta}}\right.\right\}\\
%&B_{H}=\left\{i\in[1,N]\left|Z^{H}(W_{s,N}^{(i)})<1-2^{-N^{\beta}}, \forall K\le H: Z^{K}(W_{c,N}^{(i)})>1-2^{-N^{\beta}}\right.\right\}
\end{align*}
For $H\le \G$ and $K\le \G$, define $A_{H,K}=A_H\cap B_{K}$. Note that for large $N$, $2^{-N^{\beta}} < 1-2^{-N^{\beta}}$. This implies for $i\in A_H$, we have $Z^H(W_{s,N}^{(i)})<1-2^{-N^{\beta}}$ and hence $i\in \cup_{K\le H}B_K$. Therefore, for $K \nleq H$, we have $A_{H,K}=\emptyset$. This means $\{A_{H,K}|K\le H\le \G\}$ forms a partition of $[1,N]$. Note that as $N$ increases, $\frac{|A_H|}{N}\rightarrow p_H$ and $\frac{|B_{H}|}{N}\rightarrow q_H$.\\

The encoding and decoding rules are as follows: Let $z_1^N\in\G^N$ be an outcome of the random variable $Z_1^N$ known to both the encoder and the decoder. Given $K\le H\le \G$, let $T_H$ be a transversal of $H$ in $\G$  and let $T_{K\le H}$ be a transversal of $K$ in $H$. Any element $g$ of $\G$ can be represented by $g=[g]_K+[g]_{T_{K\le H}}+[g]_{T_H}$ for unique $[g]_K\in K$, $[g]_{T_{K\le H}}\in T_{K\le H}$ and $[g]_{T_H}\in T_H$. Also note that $T_{K\le H}+T_H$ is a transversal $T_K$ of $K$ in $\G$ so that $g$ can be uniquely represented by $g=[g]_K+[g]_{T_K}$ for some $[g]_{T_K}\in T_K$ and $[g]_{T_K}$ can be uniquely represented by $[g]_{T_K}= [g]_{T_{K\le H}}+[g]_{T_H}$.

Given a source sequence $x_1^N\in\mathcal{X}^N$, the encoding rule is as follows: For $i\in[1,N]$, if $i\in A_{H,K}$ for some $K\le H\le \G$, $[a_i]_K$ is uniformly distributed over $K$ and is known to both the encoder and the decoder (and is independent from other random variables). The component $[a_i]_{T_K}$ is chosen randomly so that for $g\in \G$,
\begin{align*}
P(a_i=g)=\frac{p_{A_i|X_1^NZ_1^NA_1^{i-1}}(g|x_1^N,z_1^N,a_1^{i-1})}{p_{A_i|X_1^NZ_1^NA_1^{i-1}}([a_i]_K+T_K|x_1^N,z_1^N,a_1^{i-1})}
\end{align*}
Note that $a_1^N$ can be decomposed as $a_1^N=[v_1^N]_K+[a_1^N]_{T_{K\le H}}+[a_1^N]_{T_H}$ in which $[a_1^N]_K$ is known to the decoder. The encoder sends $[a_1^N]_{T_{K\le H}}$ to the decoder and the decoder uses the channel code to recover $[a_1^N]_{T_H}$. The decoding rule is as follows: Given $z_1^N$, $v_1^N$, $[a_1^N]_K$ and $[a_1^N]_{T_{K\le H}}$, and for $i\in A_{H,K}$, let
\begin{align*}
\hat{a}_i=\argmax_{g\in [a_i]_K+[a_i]_{T_{K\le H}}+T_H} W_{c,N}^{(i)}(z_1^N,v_1^N,\hat{a}_1^{i-1}|g)
\end{align*}
Finally, the decoder outputs $z_1^N-\hat{a}_1^NG$. Note that the rate of this code is equal to
\begin{align*}
R &=\sum_{K\le H\le \G} \frac{|A_{H,K}|}{N} \log \frac{|H|}{|K|}\\
&= \sum_{K\le H\le \G} \frac{|A_{H,K}|}{N} \log \frac{|\G|}{|K|} - \sum_{K\le H\le \G} \frac{|A_{H,K}|}{N} \log \frac{|\G|}{|H|}\\
%&= \sum_{K\le \G} \frac{|B_{K}|}{N} \log \frac{|\G|}{|K|} - \sum_{H\le \G} \frac{|A_{H}|}{N} \log \frac{|\G|}{|H|}\\
&\rightarrow \bar{I}(W_s)-\bar{I}(W_c)= I(X;U)-I(U;V)
\end{align*}

\section{Distributed Source Coding: Decoding the Sum of Variables}\label{section:KM}
\begin{figure}[!h]
\begin{minipage}[b]{0.23\textwidth}
\centering
\includegraphics[scale=.8]{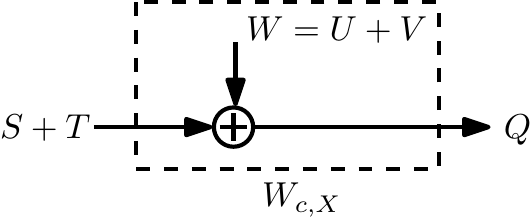}
%\caption{\small Source X: Test channel for the inner code (the channel coding component)}
\caption{\small Inner code (X).}
\label{fig:WcX_KM}
\end{minipage}
\hspace{0.5cm}
\begin{minipage}[b]{0.23\textwidth}
\centering
\includegraphics[scale=.8]{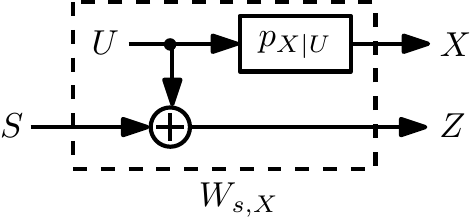}
%\caption{\small Source X: Test channel for the outer code (the source coding component)}
\caption{\small Outer code (X).}
\label{fig:WsX_KM}
\end{minipage}
\end{figure}

\begin{figure}[!h]
\begin{minipage}[b]{0.23\textwidth}
\centering
\includegraphics[scale=.8]{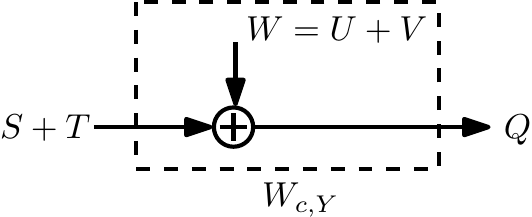}
%\caption{\small Source Y: Test channel for the inner code (the channel coding component)}
\caption{\small Inner code (Y).}
\label{fig:WcY_KM}
\end{minipage}
\hspace{0.5cm}
\begin{minipage}[b]{0.23\textwidth}
\centering
\includegraphics[scale=.8]{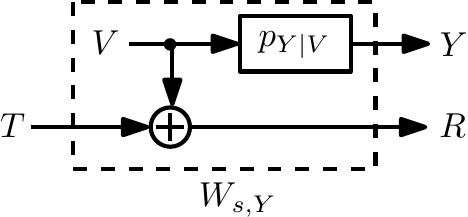}
%\caption{\small Source Y: Test channel for the outer code (the source coding component)}
\caption{\small Inner code (Y).}
\label{fig:WsY_KM}
\end{minipage}
\end{figure}

For a distributed source $(\mathcal{X}\times\mathcal{Y},p_{X,Y},d)$ let the random variables $U$ and $V$ take values from a group $\G$. Assume that $U$ and $V$ satisfy the Markov chain $U\leftrightarrow X \leftrightarrow Y \leftrightarrow V$ and assume $\mathds{E}\{d(X,Y,g(U+V))\le D\}$ for some function $g$. For $W=U+V$, we show that the following rates are achievable:
\begin{align*}
&R_1=H(W)-H(U|X),\quad R_2=H(W)-H(V|Y)
\end{align*}
for decoding $W$ at the decoder. The source $X$ employs a nested polar codes whose inner code is a good channel code for the channel $(\G,\G,W_{c,X})$ and whose outer code is a good source code for the test channel $(\G,\mathcal{X}\times\G,W_{s,X})$ where for $s,t,q,z\in G$ and $x\in \mathcal{X}$, $W_{c,X}(q|s\!+\!t)\!=\!p_W(q\!-\!s\!-\!t)$ and $W_{s,X}(x,z|s)\!=\!p_{XU}(x,z\!-\!s)$. Similarly, the source $Y$ employs a nested polar code whose inner code is a good channel code for the channel $(\G,\G,W_{c,Y})$ and whose outer code is a good source code for the test channel $(\G,\mathcal{Y}\times\G,W_{s,Y})$ where for $s,t,q,r\in G$ and $y\in \mathcal{Y}$, $W_{c,Y}(q|s\!+\!t)\!=\!p_W(q\!-\!s\!-\!t)$ and $W_{s,Y}(y,r|t)\!=\!p_{YV}(y,r\!-\!t)$. These channels are depicted in Figures \ref{fig:WcX_KM}, \ref{fig:WsX_KM}, \ref{fig:WcY_KM} and \ref{fig:WsY_KM}.

We need to show that $W_{c,X}$ is degraded with respect to $W_{s,X}$ (and $W_{c,Y}$ is degraded with respect to $W_{s,Y}$). To show this, in the definition of degradedness \cite[Definition III.1]{sahebi_nested_polar_arxiv}, we let the channel $(\G,\mathcal{X}\times \G, W)$ be such that that for $q,z\in \G$ and $x\in\mathcal{X}$, $W(q|x,z)=p_{V|X}(q-z|x)$. % Then the following proves the claim.
%\begin{align*}
%\sum_{\substack{z\in\G\\x\in\mathcal{X}}} W_{s,X}(x,z|s) W(q|x,z) &= \sum_{\substack{z\in\G\\x\in\mathcal{X}}} p_{XU}(x,z-s) p_{V|X}(q-z|x)\\
%&\stackrel{(a)}{=} \sum_{\substack{z\in\G\\x\in\mathcal{X}}} p_{XUV}(x,z-s,q-z)\\
%%&= \sum_{z'\in\G} p_{UV}(z',q-s-z')\\
%&=p_W(q-s)=W_{c,X}(q|s)
%\end{align*}
%where $(a)$ follows since the Markov chain $U\leftrightarrow X\leftrightarrow V$ holds.\\
%The rest of the analysis for this case is a little more involved compared to the previous case and will be added later.

\section{Multiple Access Channels}\label{section:MAC}
\begin{figure}[!h]
\begin{minipage}[b]{0.23\textwidth}
\centering
\includegraphics[scale=.8]{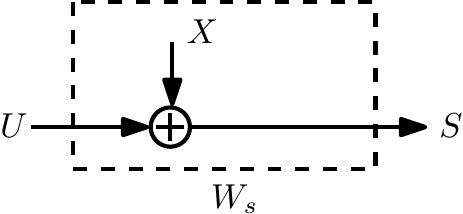}
%\caption{\small Test channel for the inner code (the source coding component)}
\caption{\small Channel for inner code.}
\label{fig:Ws_MAC}
\end{minipage}
%\hspace{0.5cm}
\begin{minipage}[b]{0.23\textwidth}
\centering
\includegraphics[scale=.8]{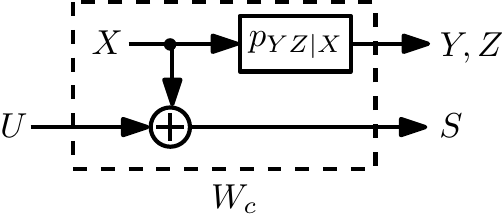}
%\caption{\small Test channel for the outer code (the channel coding component)}
\caption{\small Channel for outer code.}
\label{fig:Wc_MAC}
\end{minipage}
\end{figure}

Let the finite sets $\mathcal{X}$ and $\mathcal{Y}$ be the input alphabets of a two-user MAC and let $\mathcal{Z}$ be the output alphabet and assume the messages are independent. In order to show that nested polar codes achieve the capacity region of a MAC, it suffices to show that the rates $R_1=I(X;Z|Y)=H(X)-H(X|YZ)$ and $R_2=I(Y;Z)$ are achievable (to incorporate the time-sharing argument see Section \ref{section:discussion}). It is known from the point-to-point result \cite{sahebi_nested_polar_arxiv} that the $Y$ terminal can communicate with the decoder with rate $I(Y;Z)$ so that $y_1^N$ is available at the decoder with high probability. It remains to show that the rate $R_1$ is achievable for the $X$ terminal when $y_1^N$ is available at the decoder. Let $\G$ be an Abelian group with $|\mathcal{\G}|=|\mathcal{X}|$. Define the artificial channels $(\G,\G,W_s)$ and $(\G,\mathcal{Y}\times\mathcal{Z}\times\G,W_c)$ such that for $u,z\in \G$ and $y\in \mathcal{Y}$, $W_s(s|u)=p_X(s-u)$ and $W_c(y,z,s|u)=p_{XYZ}(s-u,y,z)$. These channels have been depicted in Figures(\ref{fig:Ws_MAC}) and (\ref{fig:Wc_MAC}).

Similarly to previous cases, one can show that the symmetric capacities of the channels are equal to $\bar{I}(W_s)=\log q-H(X)$ and $\bar{I}(W_c)=\log q-H(X|YZ)$. We employ a nested polar code in which the inner code is a good source code for the test channel $W_s$ and the outer code is a good channel code for $W_c$. The rate of this code is equal to $R=\bar{I}(W_c)-\bar{I}(W_x)=I(X;Z|Y)$. Here, we only give a sketch of the proof. First note that the channel $W_s$ is degraded with respect to $W_c$ so that the the source code is contained in the channel code.\\
For $s_1^N\in \G^N$, $y_1^N\in\mathcal{Y}^N$ and $z_1^N\in\mathcal{Z}^N$, let
\begin{align*}
&W_{s,N}^{(i)}(s_1^N,a_1^{i-1}|a_i)=\sum_{a_{i+1}^N\in \G^{N-i}} \frac{1}{q^{N-1}} W_s^N(s_1^N|a_1^NG)\\% = \sum_{a_{i+1}^N\in \G^{N-i}} \frac{1}{q^{N-1}} p_{XU}^N(x_1^N,z_1^N-a_1^NG)\\
&W_{c,N}^{(i)}(y_1^N,\!z_1^n,\!s_1^N,\!a_1^{i\!-\!1}|a_i)\!=\!\!\!\!\!\!\!\!\!\sum_{a_{i\!+\!1}^N\in \G^{N\!-\!i}} \!\!\frac{1}{q^{N-1}} W_c^N\!(y_1^N,\!z_1^N,\!s_1^N|a_1^NG)% = \sum_{a_{i+1}^N\in \G^{N-i}} \frac{1}{q^{N-1}} p_{VU}^N(v_1^N,z_1^N-a_1^NG)\\
\end{align*}

Let the random vectors $X_1^N,Y_1^N,U_1^N,V_1^N$ be distributed according to $P_{XYUV}^N$ and let $S_1^N$ be a random variable uniformly distributed over $\G^N$ which is independent of $X_1^N,Y_1^N,U_1^N,V_1^N$. Let $U_1^N=S_1^N-X_1^N$ and $A_1^N=U_1^N G^{-1}$.
%In other words, the joint distribution of the random vectors is given by
%\begin{align*}
%p_{A_1^NS_1^NU_1^NV_1^NX_1^NZ_1^N}(a_1^N,s_1^N,u_1^N,v_1^N,x_1^N,z_1^N)&=\frac{1}{q^N} p_{XUV}^N(x_1^N,u_1^N,v_1^N) \mathds{1}_{\{s_1^N=a_1^NG,u_1^n=z_1^N-a_1^NG\}}
%\end{align*}
%This implies
%\begin{align*}
%p_{A_1^NV_1^NX_1^NZ_1^N}(a_1^Nv_1^N,x_1^N,z_1^N)&=\frac{1}{q^N} p_{XUV}^N(x_1^N,z_1^N-a_1^NG,v_1^N)\\
%p_{A_1^NV_1^NZ_1^N}(a_1^N,v_1^N,z_1^N)&=\frac{1}{q^N} p_{UV}^N(z_1^N-a_1^NG,v_1^N)
%\end{align*}
%The channels $W_{s,N}^{(i)}$ and $W_{c,N}^{(i)}$ polarize in the sense defined in [???].
The encoding and decoding rules are similar to those of the point-to-point channel coding result; i.e., at the encoder, the distribution $p_{A_i|S_1^NA_1^{i-1}}$ is used for soft encoding and at the decoder, $W_{c,N}^{(i)}(y_1^N,z_1^n,s_1^N,a_1^{i-1}|a_i)$ is used in the successive cancelation decoder to decode $a_1^N$. The final decoder output is equal to $z_1^N-a_1^NG$. Note that since $y_1^N$ is known to the decoder with high probability, it can be used as the channel output for $W_c$.

\section{Computation over MAC}\label{section:Computation}
\begin{figure}[!h]
\begin{minipage}[b]{0.23\textwidth}
\centering
\includegraphics[scale=.8]{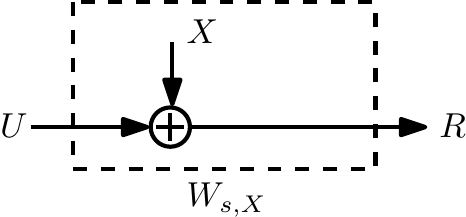}\vspace*{-.3cm}
%\caption{\small Test channel for the inner code (the source coding component)}
\caption{\small Inner code (X).}
\label{fig:WsX_Comp}
\end{minipage}
%\hspace{0.5cm}
\begin{minipage}[b]{0.23\textwidth}
\centering
\includegraphics[scale=.8]{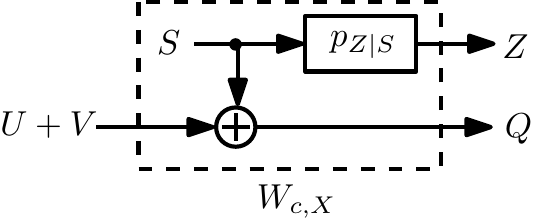}\vspace*{-.3cm}
%\caption{\small Test channel for the outer code (the channel coding component)}
\caption{\small Outer code (X).}
\label{fig:WcX_Comp}
\end{minipage}
\end{figure}

\begin{figure}[!h]
\begin{minipage}[b]{0.23\textwidth}
\centering
\includegraphics[scale=.8]{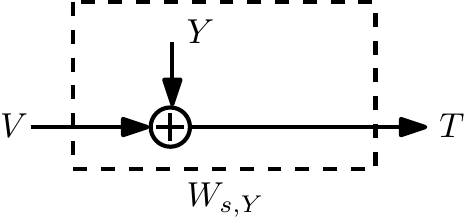}\vspace*{-.3cm}
%\caption{\small Test channel for the inner code (the source coding component)}
\caption{\small Inner code (Y).}
\label{fig:WsY_Comp}
\end{minipage}
%\hspace{0.5cm}
\begin{minipage}[b]{0.23\textwidth}
\centering
\includegraphics[scale=.8]{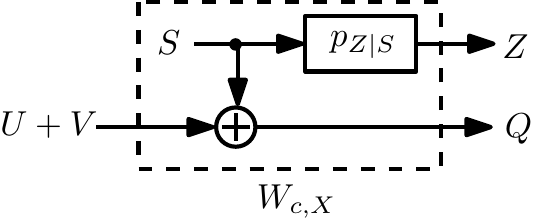}\vspace*{-.3cm}
%\caption{\small Test channel for the outer code (the channel coding component)}
\caption{\small Outer code (X).}
\label{fig:WcY_Comp}
\end{minipage}
\end{figure}

In this section, we consider a simple computation problem over a MAC with input alphabets $\mathcal{X}$, $\mathcal{Y}$ and output alphabet $\mathcal{Z}$. The two input terminals of a MAC, $X$ and $Y$ are trying to communicate with a centralized decoder which is interested in the sum of the inputs $S=X+Y$ where $+$ is summation over a group $\G$. We show that the rate $R=\min(H(X),H(Y))-H(S|Z)$ is achievable using polar codes.
The terminal $X$ employs a nested polar code whose inner code is a good source code for the test channel $(\G,\G,W_{s,X})$ and whose outer code is a good channel code for the channel $(\G,\mathcal{Z}\times\G,W_{c,X})$ where for $u,v,r,z\in \G$ and $z\in \mathcal{Z}$, $W_{s,X}(r|u)\!=\!p_X(r\!-\!u)$ and $W_{c,X}(z,q|u\!+\!v)\!=\!p_{SZ}(q\!-\!u\!-\!v,z)$. Similarly, the terminal $Y$ employs a nested polar code whose inner code is a good source code for the test channel $(\G,\G,W_{s,Y})$ and whose outer code is a good channel code for the channel $(\G,\mathcal{Z}\times\G,W_{c,Y})$ where for $u,v,t,z\in \G$ and $z\in \mathcal{Z}$, $W_{s,Y}(t|v)\!=\!p_Y(t\!-\!v)$ and $W_{c,Y}(z,q|u\!+\!v)\!=\!p_{SZ}(q\!-\!u\!-\!v,z)$. Note that the two terminals use the same channel code. These channels are depicted in Figures \ref{fig:WsX_Comp}, \ref{fig:WcX_Comp}, \ref{fig:WcY_Comp} and \ref{fig:WsY_Comp}.

\vspace*{-.3cm}
Similarly to previous cases, one can show that the symmetric capacities of the channels are equal to $\bar{I}(W_s)=\log q-H(X)$ and $\bar{I}(W_c)=\log q-H(X|YZ)$. We employ a nested polar code in which the inner code is a good source code for both test channels $W_{s,X}$ and $W_{s,Y}$ and the outer code is a good channel code for $W_{c,X}=W_{c,Y}$. The rate of this code is equal to $R=\bar{I}(W_{c,X})-\max(\bar{I}(W_{s,X}),\bar{I}(W_{s,Y}))=\min(H(X),H(Y))-H(S|Z)$. It is worth noting that it can be shown that the intersection of the two source codes is contained in the common channel code.% This has been shown in [???].

\section{The Broadcast Channel}\label{section:BC}
\begin{figure}[!h]
\begin{minipage}[b]{0.23\textwidth}
\centering
\includegraphics[scale=.8]{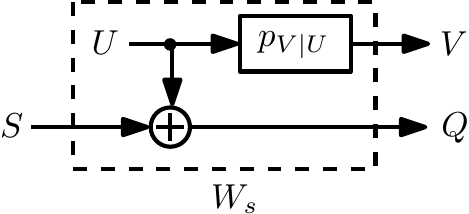}\vspace*{-.3cm}
\caption{\small Channel for inner code}
\label{fig:Ws_BC}
\end{minipage}
\hspace{0.5cm}
\begin{minipage}[b]{0.23\textwidth}
\centering
\includegraphics[scale=.8]{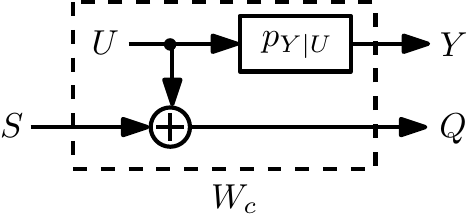}\vspace*{-.3cm}
\caption{\small Channel for outer code}
\label{fig:Wc_BC}
\end{minipage}
\end{figure}

In this section, we consider a broadcast channel $(\mathcal{X},\mathcal{Y}\times\mathcal{Z},W,w)$ when $\mathcal{X}=\G$ for some arbitrary Abelian group $\G$. Let $X$ be a random variable over $\mathcal{X}$ such that $\mathds{E}\{w(X)\}\le D$ and let $Y,Z$ be the corresponding channel outputs. Let $U,V$ be random variable over $\G$ satisfying the Markov chain $UV\leftrightarrow X \leftrightarrow YZ$ such that there exists a function $g:\G^2\rightarrow \mathcal{X}$ with $g(U,V)=X$. We show that the following rates are achievable
\begin{align*}
&R_1\!=\!I(U;Y)\!-\!I(U;V)\!=\!H(U|Y)\!-\!H(U|V),R_2\!=\!I(V;Z)
\end{align*}
if the Markov chain $U\leftrightarrow X\leftrightarrow V$ holds in addition to the Markov chain above needed for Marton's bound. Note that the $Z$ terminal can use a point-to-point channel code to achieve the desired rate. It remains to show that the rate $R_2$ is achievable when $v_1^N$ is available at the encoder. Define the artificial channels $(\G,\G^2,W_s)$ and $(\G,\mathcal{Y}\times\G,W_c)$ such that for $s,v,z\in \G$ and $y\in \mathcal{Y}$, $W_s(v,z|s)=p_{UV}(z-s,v)$ and $W_c(y,z|s)=p_{UY}(z-s,y)$. These channels have been depicted in Figures(\ref{fig:Ws_BC}) and (\ref{fig:Wc_BC}).\vspace*{-.3cm}
\vspace*{-.3cm}
Similarly to previous cases, one can show that the symmetric capacities of the channels are equal to $\bar{I}(W_s)=\log q-H(U|V)$ and $\bar{I}(W_c)=\log q-H(U|Y)$. Note that to guarantee that $W_s$ is degraded with respect to $W_c$, we need an additional condition on the auxiliary random variables. It suffices to assume that the Markov chain $U\leftrightarrow X\leftrightarrow V$ holds.% It is shown in [???] that this assumption does not reduce the rate region (I'm not 100\% sure this is true, can we discuss this?).
%
%Let $\mathds{C}_1\subset\mathcal{U}^n$ and $\mathds{C}_2\subset\mathcal{V}^n$ be a code in Marton's ensemble. For a pair of messages $u_1^N,v_1^N$, $x_1^N=g(u_1^N,v_1^N)$ is sent to the channel. If the function $g(\cdot,\cdot)$ is not injective, there exist $u,\tilde{u}\in\mathcal{U}$ and $v,\tilde{v}\in\mathcal{V}$ such that $(u,v)\ne (\tilde{u},\tilde{v})$ and $g(u,v)=g(\tilde{u},\tilde{v})$.

We employ a nested polar code in which the inner code is a good source code for the test channel $W_s$ and the outer code is a good channel code for $W_c$. The rate of this code is equal to $R=\bar{I}(W_c)-\bar{I}(W_x)=I(U;Y)-I(U;V)$.

\section{Multiple Description Coding}\label{section:MD}
\begin{figure}[!h]
\begin{minipage}[b]{0.23\textwidth}
\centering
\includegraphics[scale=.8]{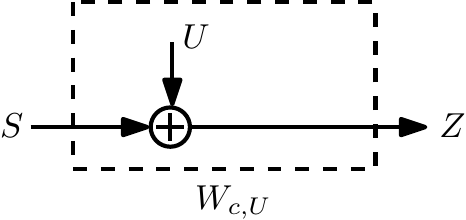}\vspace*{-.3cm}
%\caption{\small Test channel for the inner code (the source coding component)}
\caption{\small Inner code (X).}
\label{fig:WcU_MD}
\end{minipage}
%\hspace{0.5cm}
\begin{minipage}[b]{0.23\textwidth}
\centering
\includegraphics[scale=.8]{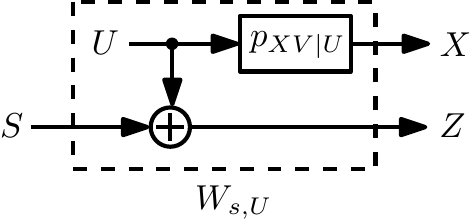}\vspace*{-.3cm}
%\caption{\small Test channel for the outer code (the channel coding component)}
\caption{\small Outer code (X).}
\label{fig:WsU_MD}
\end{minipage}
\end{figure}

\begin{figure}[!h]
\begin{minipage}[b]{0.23\textwidth}
\centering
\includegraphics[scale=.8]{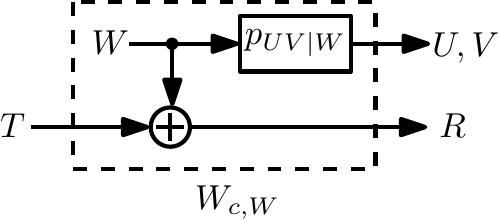}\vspace*{-.3cm}
%\caption{\small Test channel for the inner code (the source coding component)}
\caption{\small Inner code (Y).}
\label{fig:WcW_MD}
\end{minipage}
%\hspace{0.5cm}
\begin{minipage}[b]{0.23\textwidth}
\centering
\includegraphics[scale=.8]{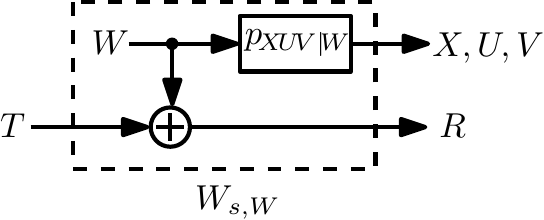}\vspace*{-.3cm}
%\caption{\small Test channel for the outer code (the channel coding component)}
\caption{\small Outer code (Y).}
\label{fig:WsW_MD}
\end{minipage}
\end{figure}

Consider a multiple description problem in which a source $X$ is to be reconstructed at three terminals $U$, $V$ and $W$. There are two encoders and three decoders. Terminals $U$ and $V$ have access to the output of their corresponding encoders and terminal $W$ has access to the output of both encoders. The goal is to find all achievable tuples $(R_1,R_2,D_1,D_2,D_3)$ where $R_1$ and $R_2$ are the rates of encoders $U$ and $V$ respectively and $D_1$, $D_2$ and $D_3$ are the distortion levels corresponding to decoders $U$, $V$ and $W$ respectively. $D_1$, $D_2$ and $D_3$ are measured as the average of distortion measures $d_1(\cdot,\cdot)$, $d_2(\cdot,\cdot)$ and $d_3(\cdot,\cdot)$ respectively. Let $U$, $V$ and $W$ be random variables such that $\mathds{E}\{d_1(X,U)\}\le D_1$, $\mathds{E}\{d_2(X,V)\}\le D_2$ and $\mathds{E}\{d_3(X,W)\}\le D_3$. We show that the tuple $(R_1,R_2,D_1,D_2,D_3)$ is achievable if $R_1\ge I(X;U)$, $R_2\ge I(X;V)$ and $R_1+R_2\ge I(X;UVW)+I(U;V)$. It suffices to show that the rates $R_1=I(X;UVW)-I(X;V)+I(U;V)$, $R_2=I(X;V)$ are achievable. The point-to-point source coding result implies that with $R_2=I(X;V)$ we can have $v_1^N$ at the output of the second decoder with high probability. To achieve the rate $R_1$ when $v_1^N$ is available, first we note that $R_1=H(U)-H(U|VX)+H(W|UV)-H(W|UVX)$. We use a code with rate $R_{11}=H(U)-H(U|VX)$ for sending $U$ and another code $R_{12}=H(W|UV)-H(W|UVX)$ for sending $W$. The corresponding channels are depicted in Figures  \ref{fig:WcU_MD}, \ref{fig:WsU_MD}, \ref{fig:WcW_MD} and \ref{fig:WsW_MD}.

\vspace*{-.5cm}
\vspace*{-.3cm}
%Similarly to previous cases, one can show that the symmetric capacities of the channels are equal to $\bar{I}(W_s)=\log q-H(X)$ and $\bar{I}(W_c)=\log q-H(X|YZ)$. We employ a nested polar code in which the inner code is a good source code for both test channels $W_{s,X}$ and $W_{s,Y}$ and the outer code is a good channel code for $W_{c,X}=W_{c<Y}$. The rate of this code is equal to $R=\bar{I}(W_{c,X})-\max(\bar{I}(W_{s,X}),\bar{I}(W_{s,Y}))=\min(H(X),H(Y))-H(S|Z)$. %It is worth noting that a first step in the proof is to show that the intersection of the two source codes is contained in the common channel code. This has been shown in [???].

%\section{The Interference Channel}

\section{Other Problems and Discussion}\label{section:discussion}
In this paper, we studied the main multi-terminal communication problems in their simplest forms (e.g., no time sharing etc.). The approach of this paper can be extended to the more general formulations and to other similar problems. The approach presented in this paper can also be extended to multiple user (more than two) cases in a straightforward fashion. We briefly discuss examples of such extensions. First, consider the Berger-Tung rate region for the distributed source coding problem and let $Q$ be the time-sharing random variable. We show that the rates $R_1=I(X;U|VQ)$ and $R_2=I(Y;V|Q)$ are achievable for this problem. To achieve these rates, note that $R_1=I(U;XQ)-I(U;VQ)$ and $R_2=I(V;YQ)-I(V;Q)$ and design an inner polar code of rate $I(U;VQ)$ and an outer polar code of rate $I(U;XQ)$ for the source $X$, and an inner polar code of rate $I(V;Q)$ and an outer polar code of rate $I(V;YQ)$ for the source $Y$ using some suitably defined channels. Let's denote the channel depicted in Figure \ref{fig:Wc_RD} symbolically by $\mathcal{W}(U\rightarrow V)$ for generic random variables $U$ and $V$. Then, these channels are given by $\mathcal{W}(U\rightarrow V,Q)$, $\mathcal{W}(U\rightarrow X,Q)$, $\mathcal{W}(V\rightarrow Q)$ and $\mathcal{W}(V\rightarrow Y,Q)$ respectively. %illustrated in Figures \ref{fig:Wc_DSC_Q} and \ref{fig:Ws_DSC_Q} for the terminal $X$.%, \ref{Wc_RD_Q} and \ref{Ws_RD_Q}
%\begin{figure}[!h]
%\begin{minipage}[b]{0.23\textwidth}
%\centering
%\includegraphics[scale=.8]{Wc_DSC_Q.pdf}
%\caption{\small Channel for inner code (X).}
%\label{fig:Wc_DSC_Q}
%\end{minipage}
%\hspace{0.5cm}
%\begin{minipage}[b]{0.23\textwidth}
%\centering
%\includegraphics[scale=.8]{Ws_DSC_Q.pdf}
%\caption{\small Channel for outer code (X).}
%\label{fig:Ws_DSC_Q}
%\end{minipage}
%\end{figure}

%\begin{figure}[!h]
%\begin{minipage}[b]{0.23\textwidth}
%\centering
%\includegraphics[scale=.8]{Wc_RD_Q.pdf}
%\caption{\small Channel for inner code (Y).}
%\label{fig:Wc_RD_Q}
%\end{minipage}
%\hspace{0.5cm}
%\begin{minipage}[b]{0.23\textwidth}
%\centering
%\includegraphics[scale=.8]{Ws_RD_Q.pdf}
%\caption{\small Channel for outer code (Y).}
%\label{fig:Ws_RD_Q}
%\end{minipage}
%\end{figure}
Next, we consider the multiple description coding problem and show that the rates $R_1=I(X;VT)$ and $R_2=I(X;UVW|T)+2I(X;T)+I(U;V|T)-I(X;VT)$ are achievable for some random variable $T$. To achieve these rates, we note that $R_1=R_{11}+R_{12}+R_{13}$ and $R_2=R_{21}+R_{22}$ where $R_{11}=H(U|T)-H(U|XVT)$, $R_{12}=H(W|UVT)-H(W|XUVT)$, $R_{13}=H(T)-H(T|X)$, $R_{21}=H(V|T)-H(V|XT)$ and $R_{22}=H(T)-H(T|X)$. We design a nested polar codes for each of these rates similarly to the other examples presented in the paper.

Finally, consider a 3-user MAC with inputs $W$, $X$ and $Y$ and output $Z$. We have seen in Section \ref{section:MAC} that with rates $R_X=I(X;YZ)$ and $R_Y=I(Y;Z)$, we can have access to $x_1^N$ and $y_1^N$ at the decoder with high probability. The channels $\mathcal{W}(W\rightarrow 0)$ and $\mathcal{W}(V\rightarrow X,Y,Z)$ can be used to design a nested polar code of rate $R_W=I(W;Z|XY)$ for terminal $W$.
%The channels $W_{s,W}$ and $W_{c,W}$ depicted in Figures \ref{fig:Ws_3MAC} and \ref{fig:Wc_3MAC} can be used to design a nested polar code of rate $R_W=I(W;Z|XY)$ for terminal $W$.
%\begin{figure}[!h]
%\begin{minipage}[b]{0.23\textwidth}
%\centering
%\includegraphics[scale=.8]{Ws_3MAC.pdf}
%\caption{\small Channel for inner code.}
%\label{fig:Ws_3MAC}
%\end{minipage}
%\hspace{0.5cm}
%\begin{minipage}[b]{0.23\textwidth}
%\centering
%\includegraphics[scale=.8]{Wc_3MAC.pdf}
%\caption{\small Channel for outer code.}
%\label{fig:Wc_3MAC}
%\end{minipage}
%\end{figure}

\bibliographystyle{IEEEtran}
\bibliography{IEEEabrv,ariabib}
\end{document}